# OH⁻-Enhanced Alkaline Hydrogen Evolution Reaction at the Au(111) Electrode

Er-Fei Zhen,[1] Tao-Qi Zong,[1] Lang Qin,[1] Bai-Quan Zhu,[1] Jun Cai,[1] Marko M. Melander,[2] and Yan-Xia Chen*,[1]

[1]Hefei National Research Center for Physical Sciences at Microscale, Department of Chemical Physics, University of Science and Technology of China, Hefei, 230026, China
[2]Department of Chemistry, Nanoscience Center, University of Jyväskylä, P.O. Box 35 (YN), FI-40014 Jyväskylä, Finland

*Corresponding Author Email: yachen@ustc.edu.cn

**Abstract**: The hydrogen evolution reaction (HER) in alkaline media suffers from sluggish kinetics but the origin of the pH-dependent activity remains debated. This study investigates the role of hydroxide ions (OH⁻) in enhancing the alkaline HER at Au(111) by systematically varying the pH and the NaOH concentration both with and without fixing the total Na concentration. Contrary to conventional cation-centric models of alkaline HER, we demonstrate a notable anion effect by showing that the HER activity increases monotonically with pH and OH⁻ concentration, even at extremely high NaOH concentrations (up to 9 M). Tafel slopes decrease from ∼181 mV/dec at pH ≈10 to ∼124 mV/dec at pH ≈13 and to ∼111 mV/dec for the case with 9 M NaOH, indicating accelerated kinetics. Infrared spectroscopy reveals that interfacial OH⁻ strengthens the hydrogen-bond network, which is expected to lower the activation energy for the Volmer step, the rate-determining step of HER on Au(111). Hence, OH⁻ enhances alkaline HER kinetics by strengthening the hydrogen bond network and its connectivity at the electrochemical interface; this allows us to propose a unified mechanism for the electrolyte effects on alkaline HER where structure-making ions (Li⁺, K⁺, and OH⁻) improve the reaction kinetics by optimizing the interfacial hydrogen bond network. These findings and insights establish OH⁻ as a critical promoter of alkaline HER, offering new strategies for designing high-performance alkaline electrolyzers through interfacial electrolyte engineering.

## 1. Introduction

Hydrogen electrocatalysis has emerged as a cornerstone for sustainable energy systems, offering a pathway to decarbonize industries and to enable large-scale hydrogen production via water electrolysis.[1, 2] As a clean energy carrier, hydrogen boasts a high energy density and zero carbon emissions, positioning it as a critical enabler of global carbon neutrality goals.[3, 4] Among hydrogen production technologies, alkaline water electrolysis stands out for its compatibility with non-precious metal catalysts and cost-effective infrastructure.[5-7] However, the hydrogen evolution reaction (HER) in alkaline media suffers from sluggish kinetics. For instance, even platinum (Pt), the benchmark HER catalyst, exhibits a reduction in HER activity of more than two orders of magnitude in alkaline environments compared to that in acidic conditions.[8] The hot debate on the origin(s) of pH-dependent HER activity has continued for the past two decades[8-15] and the incomplete understand on pH effects still hinders the development of alkaline water electrolyzers.

On the metal side, researchers have made substantial progress in optimizing the electrocatalysts to overcome the sluggish kinetics of alkaline HER. Significant advancements have been achieved on metal-based catalysts through alloying,[16-18] heterostructure engineering,[19-21] strain engineering[22, 23] and doping,[24-26] etc. These developments establish crucial material design principles and activity descriptors that enable precise regulation of adsorption energies of key reaction intermediates, particularly adsorbed hydrogen ($H_{ad}$) and hydroxyl ($OH_{ad}$).[27-30] Such systematic studies bridge the gap between catalyst design and alkaline HER/HOR performance, advancing scalable green hydrogen production.

On the electrolyte side, the electrolyte effects, particularly cation effects, have gradually gained traction as a novel lever to tune HER kinetics.[31, 32] Intriguingly, the cation identity dictates activity trends: HER rates on Pt-group metals in alkaline medium are higher in electrolytes that contain cations with a smaller ionic radius ($Li^+ > Na^+ > K^+ > Rb^+ > Cs^+$). On coinage metals (e.g., Au, Ag), some studies have reported[33-35] that alkaline HER activity exhibits a reversed trend and is more active in the presence of larger cations ($Li^+ < Na^+ < K^+ < Rb^+ < Cs^+$). In contrast, a recent study[36] found that HER activity on Au increases in the order of $Li^+ > Na^+ > K^+ > Rb^+ > Cs^+$ in acid, neutral and basic solutions. Furthermore, several theories,[31] such as hydrogen adsorption energy modulation,[35] solvent reorganization effects,[37] hard-soft acid-base theory,[38] and interfacial hydrogen-bond connectivity and blocking of hydrogen transport to the electrode surface,[39] have been put forward to rationalize these cation effects.

Beyond the cation identity, the cation concentration also plays a critical role.[33, 40, 41] Pioneering work by Marc Koper's group revealed that increasing cation concentration (e.g., $Na^+$) on Au electrodes initially enhances HER activity by strengthening the interfacial electric field, which facilitates water orientation and stabilizes transition states of the Volmer step.[40] However, excessively high cation concentrations induce ion crowding, leading to the formation of a rigid Helmholtz layer that impedes proton transport and suppresses activity.[39,40] This non-monotonic behavior highlights the delicate balance between cation-induced electric field effects and steric hindrance, emphasizing the uniqueness of interfacial cation structuring in governing HER dynamics.

Especially the interfacial hydrogen-bond connectivity and hydrogen transport blocking lead to a question: at what cation concentration does the effect on the HER change from promotion to inhibition? Inspired by work from the Koper group[40] who studied the HER up to ~0.15 M $Na^+$ @ pH = 13, we address this question by studying the HER on Au(111) at higher NaOH concentrations, and observe that the HER activity and kinetics improve *monotonously* as the NaOH concentration is increased above 1 M; even with NaOH concentration up to 9 M the blocking effect of cations is not observed. This result cannot be explained by the prevalent cation-centric view of electrolyte effects arising from the hydrogen-bond connectivity and hydrogen transport blocking, which motivated us to explore the underlying reasons behind our observation in detail. In particular, we focus on the role of the anions ($OH^-$) and consider whether higher $OH^-$ concentrations could promote the HER kinetics. While $OH^-$ may influence local pH, alter intermediate adsorption energies, or participate in hydrogen-bond network, their specific contributions remain ambiguous. In this study, we employ Au(111) as a model electrode, and HER as a model reaction, to unravel the interplay among the interfacial cation, $OH^-$, and HER activity.

## 2. Experimental section

All electrolyte solutions were prepared using Millipore water (18.2 MΩ·cm). 0.05 M $H_2SO_4$ or x M NaOH solutions were used as a supporting electrolyte and they were prepared using $H_2SO_4$ (70% ~ 72%, Aladdin) or NaOH (99.99%, Merck, 32% solution). NaOH is used with further purification in the electrolysis experiments. The purification of NaOH was performed as follows: 500 ml 32% solution of NaOH electrolyte was prepared. Two pieces of Au foil (Puratronic 99.999%, 0.25 mm; TANAKA Precious Metals) were used as working and counter electrodes. The surface area of the working electrode was 2 cm$^2$. The electrolysis was carried out in a polypropene bottle under stirring of the electrolyte at 500 rpm and under purging of the electrolyte with Ar (5 N) at 10 sccm. The applied potential was −2 V *vs*. the Au foil. The electrodes were removed from the electrolyte under applied bias after ≈30 hours at the current of 420 mA.

A conventional three-electrode cell was used in the electrochemical measurements. The electrolyte solutions were de-aerated with purified Ar gas (5 N) for 20 min before each experiment. During the measurements, the electrolyte was constantly purged with Ar. Au(111) was prepared according to Clavier's method [42] and used as the working electrode (WE) while a saturated calomel electrode (SCE), connected to the cell through a long Luggin capillary, was used as the reference electrode (RE), and an Au wire was used as the counter electrode (CE). Before each experiment, the Au(111) electrode was flame-annealed followed by quenching in ultrapure water. During the transfer of the WE, the electrode surface was protected with ultrapure water to prevent contamination by airborne impurities. The measurements of the $j - E$ curves were carried out in the hanging meniscus rotating disk electrode (HMRDE) configuration with a rotation speed of 2500 rpm. Unless specifically mentioned, the meniscus height is carefully adjusted to make sure that only the sites from the (111) plane (no edge sites from the sphere) are involved in the reaction (Figure. S1c). Ohmic compensation, i.e. IR corrections, of 95% was used for all measurements.[43] The currents were normalized to the geometric surface areas of the WE to obtain the current density. All measurements are done at room temperature, all the potentials quoted in this study are against RHE unless otherwise mentioned. To make sure the results are free of artefact such as contaminants, change of meniscus height and so on, control experiments have been carried out systematically.[44]

In the ATR-FTIRS studies, a Au thin film deposited on a hemi-cylindrical Si prism (geometric area: 1.76 cm$^2$) with a thickness of 50 nm by electroless deposition method was used as the working electrode (WE).[45] An Au wire and an Ag/AgCl electrode (with saturated KCl solution) were utilized as the counter electrode (CE) and reference electrode (RE), respectively. Before each IR measurement, potential sweeping was performed in 0.1 M HClO$_4$ solution over the potential range from 0.2 to 1.6 V at a scan rate of 100 mV/s, to clean the electrode surface and ensure reproducibility. The configuration of the spectro-electrochemical cell used in the study was similar to that described in Ref. [46] The *in-situ* ATR-FTIRS experiments were carried out using a Nicolet iS50 spectrometer equipped with an MCT detector. The IR spectra were recorded simultaneously with the potential sweeping at a scan rate of 5 mV/s. Each spectrum was obtained by integrating 16 interferograms at a resolution of 4 cm$^{-1}$. All spectra are presented in absorbance units, *i.e.*, -log($R/R_0$), with $R$ and $R_0$ representing the reflected radiation intensities of the sample and reference spectra, respectively. The potentials given in the figures are against reversible hydrogen electrode (RHE), unless specially addressed. For comparison, spectra of the bulk solution with x M NaOH (x = 0 to 10) were measured under both the transmission mode and ATR mode, with the spectrum of CaF$_2$|Air|CaF$_2$ interface and Si/Air interface taken as the background, respectively.

## 3. Results and discussion

Before describing and discussing the experimental results, we will first briefly describe the mechanisms and rate equation for HER at Au(111). In alkaline medium, water molecules are the hydrogen donors and the overall HER mechanism is

$$2H_2O + 2e^- \rightleftharpoons H_2 + 2OH^- \tag{R1}$$

with the following elementary steps:

$$\text{Volmer step: } H_2O + * + e^- \rightleftharpoons H_{ad} + OH^- \tag{R2}$$

$$\text{Heyrovsky step: } H_2O + H_{ad} + e^- \rightleftharpoons H_2 + OH^- + * \tag{R3}$$

Our previous work [47] confirms that the Volmer reaction is the rate-determining step for HER at Au(111) and that the reverse reaction of the Volmer step negligible. In this case, the overall HER current density can be expressed as

$$j_{\text{alka}} = 2j_{2f} = -2e_0 n_M \frac{c_{H_2O}^{RP}}{c_{H_2O}^0} A \exp\left(-\frac{\Delta G_{\text{alkaV}}^{\neq}(E_{\text{SHE}})}{RT}\right) \tag{1}$$

where $j_{2f}$ specifically refers to the current density of the forward alkaline Volmer step (R2) and $A$ is the pre-exponential factor; the superscript RP denotes the reaction plane, $c_{H_2O}^{RP}$ represent the concentration of protons and water at the RP, $n_M$ is the number density of Au atoms on Au(111) electrode, $R$ is the gas constant, and $T$ is the absolute temperature. $e_0$ is the elementary charge and $c_{H_2O}^0$ the concentration of water under standard conditions. Because water molecules are uncharged and their concentration is higher than that of other solution-phase species, we assume that $c_{H_2O}^{RP} \approx c_{H_2O}^{\text{bulk}}$ when the electrolyte concentration is not so high. $\Delta G_{\text{alkaV}}^{\neq}(E_{\text{SHE}})$ is the activation energy of the Volmer step of water discharge which is a function of the applied electrode potential ($E_{\text{SHE}}$), on the standard hydrogen electrode scale (SHE). Assuming that the reaction barrier follows to the Brønsted-Evans-Polanyi relation[48] results in

$$\Delta G^{\neq}_{\text{alkaV}}(E_{\text{SHE}}) = \Delta G^{\neq,0}_{\text{alkaV}} + \alpha_{\text{alkaV}}F(E_{\text{SHE}} - \phi^{\text{RP}}_{\text{alka}} - E^0_{\text{SHE,alkaV}}) \qquad (2)$$

where $\Delta G^{0,\neq}_{\text{alkaV}}$ is the activation energy of the Volmer reaction when the overall HER is under standard equilibrium conditions. $\alpha_{\text{alkaV}}$ is the charge transfer coefficient of R2. The applied electrode potential ($E_{\text{SHE}}$) can be written as $E_{\text{SHE}} = \phi_M - \phi_S - \mu^M_e/F - E^{\text{SHE}}_{\text{abs}}$,[49] where $\phi_M$ and $\phi_S$ are the inner electric potential of the metal electrode and the solution, $\mu^M_e$ is the chemical potential of electrons in the metal electrode, which can be estimated from DFT calculations,[50, 51] and $E^{\text{SHE}}_{\text{abs}}$ is the absolute potential of the standard hydrogen electrode. $\phi^{\text{RP}}_{\text{alka}}$ is the electric potential at the RP in alkaline media. $E^0_{\text{SHE,alkaV}}$ is the equilibrium potential of the Volmer step in alkaline conditions when the overall HER reaction is under standard equilibrium conditions.

Figure 1a shows that the polarization curves ($j - E$) for HER at Au(111) under the same electrode potential (*vs.* reversible hydrogen electrode, RHE) increase with increasing pH from 10 to 13, consistent with the trend reported by Koper et al.[40] According to the traditional interpretations[40], the enhanced HER current at higher pH arises from the positive shift in the potential of zero free charge (PZFC) by ~59 mV per pH unit under the RHE scale. This shift leads to a more negatively charged Au(111) surface at identical HER potentials on the RHE scale in higher pH solutions and thereby increased accumulation of hydrated Na⁺ ions in the electrical double layer (EDL). The cations are expected to stabilize the transition state where water molecules dissociate to an adsorbed hydrogen and solvated $OH^-$, which in turn would lead to improved HER kinetics as the local cation concentration increases.[40] As the cation concentration and pH are both controlled by the NaOH concentration, the HER current increases with pH and the cation concentration. Although this approach allows treating the interfacial cation concentration as a descriptor to explain the cation concentration and identity effects on HER for both Au and Pt electrodes in qualitative and self-consistent manner,[33] it cannot guarantee or exclude contributions from other possible factors which accompany the changes in solution pH.

To eliminate the influence of the pH-dependent shift on the PZFC, we converted the potential scale to SHE. As shown in Figure 1b, at the same $E_{\text{SHE}}$ ($E_{\text{SHE}} < -1.35$ V$_{\text{SHE}}$), HER currents still increase with pH when the total cation concentration is fixed. At −1.6 V$_{\text{SHE}}$, the HER current at pH ≈ 13 is ~3.7 times higher than at pH ≈10. This observation is counterintuitive for two reasons. First, the rate-determining Volmer step for alkaline HER at Au(111) involves water dissociation (R2). If Frumkin-Butler-Volmer theory (Eqs.1 and 2) should hold for HER at Au(111) and if the pH-induced changes in $\Delta G_{\text{alkaV}}^{0,\neq}$ and $\alpha_{\text{alkaV}}$ are negligible, the HER rate should be pH-independent under the SHE scale. Second, under the SHE scale, the PZFC of Au(111)/electrolyte interface does not change with solution pH:[52] this fixed PZFC ensures that the surface charge density of Au(111) at the same $E_{\text{SHE}}$ does not depend on the pH. Thus, the hydrated Na$^+$ concentration (Figure 2a and 2c) and electric potential (Figure 2b and 2d) within the EDL under HER conditions should also remain unchanged at the same $E_{\text{SHE}}$. This in turn indicates that any cation effects (promotion or inhibition) at a given potential on the SHE scale should remain unchanged regardless of the pH of bulk solution. However, the results in Figure 1 clearly show that the HER current density depends sensitively on the pH and the $\text{OH}^-$ concentration even when the cation concentration remains constant and that the HER current density at a given potential becomes higher as the pH is increased.

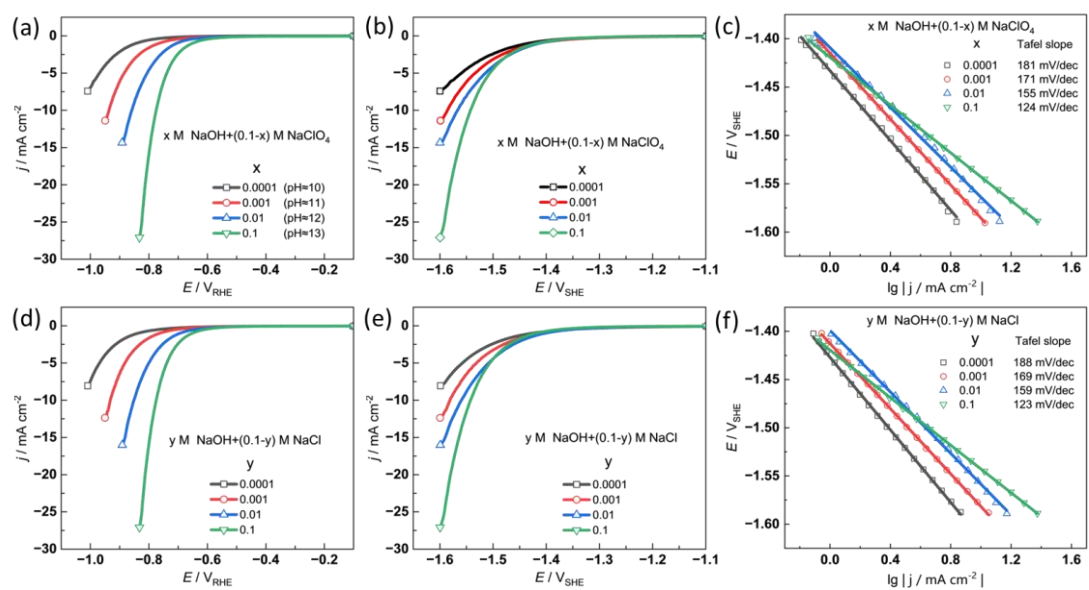

Figure 1 Polarization curves ($j - E$) and the corresponding Tafel plots for HER at Au(111) surface recorded at 10 mV/s and 2500 rpm in Ar-saturated (a-c) x M NaOH+(0.1-x) M NaClO$_4$ or (d-f) y M NaOH+(0.1-y) M NaCl [x or y =0.0001 (pH≈10),0.001 (pH≈11),0.01 (pH≈12), 0.1 (pH≈13)] solutions on SHE scale; All potentials were corrected for the iR-drop effects.

Besides a notable increase in the HER activity with an increased pH, also the Tafel slope decreases (Figure 1c), from ~181 mV/dec at pH ≈10 to ~124 mV/dec at pH ≈13. This provides direct evidence that the HER kinetics are enhanced by the higher pH. Notably, all variables except pH and solution composition were rigorously controlled, and experiments were repeated three times to exclude experimental errors or artifacts.[53] Initially, we speculated that perchlorate ions ($ClO_4^-$) in the electrolyte might suppress HER activity in solutions of lower pH when more $NaClO_4$ is added. Although the negatively charged Au(111) surface under HER conditions repels $ClO_4^-$, its "structure-breaker" characteristics[54] could disrupt hydrogen-bond connectivity in the outer EDL and consequently reduce HER activity. To test this hypothesis, we replaced $ClO_4^-$ with chloride ions ($Cl^-$) and repeated the alkaline HER experiments. As shown in Figure 1d–1f and 1a–1c, the pH-dependent trends in $j-E$ curves and Tafel slopes are nearly identical in the presence of $Cl^-$ and $ClO_4^-$, respectively. This indicates that the electrolyte counter anion ($Cl^-$ and $ClO_4^-$) has a minimal contribution to HER activity.

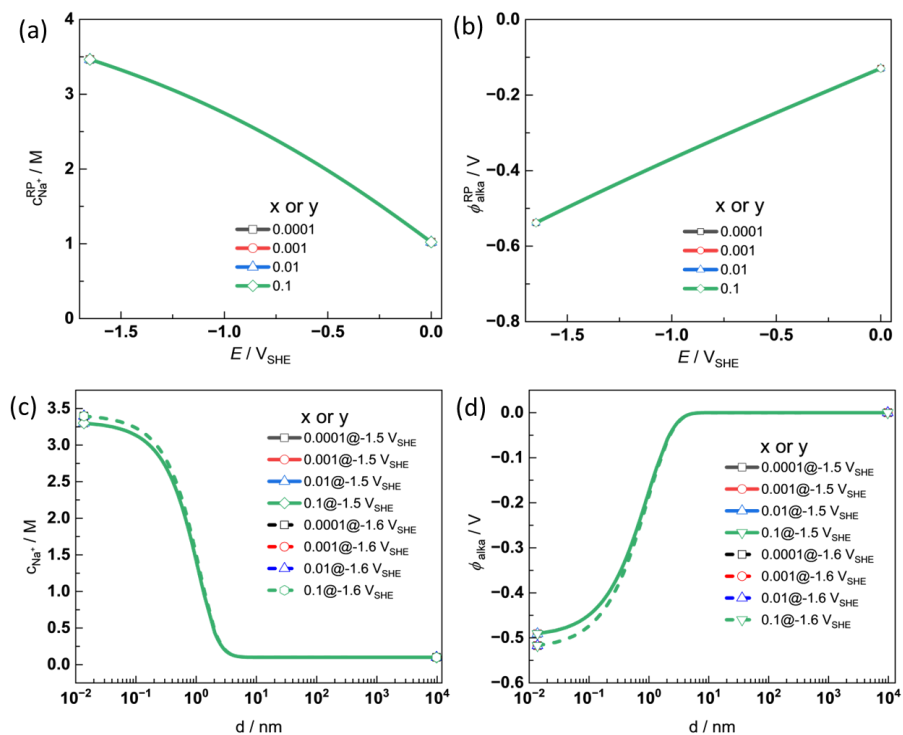

Figure 2 (a) The concentration of hydrated $Na^+$ at the RP, $c_{Na^+}^{RP}$ and (b) the electric potential at the RP, $\phi_{alka}^{RP}$, as a function of the applied electrode potential for Au(111) in x M NaOH+(0.1-x) M $NaClO_4$ or y M NaOH+(0.1-y) M NaCl [x or y = 0.0001,0.001,0.01,0.1] solutions, respectively. The RP is assumed to be just the outer Helmholtz plane (0.25 nm to Au(111) surface).[47] (c) $c_{Na^+}$ and (d) $\phi_{alka}$ as a function of the distance from Au(111) surface at $E = -1.50\ V_{SHE}$ and $E = -1.60\ V_{SHE}$ during HER, respectively.

Since the impact of $Na^+$ concentration and influence of counter-anions ($ClO_4^-$ or $Cl^-$) on the HER current density and kinetics can be excluded through the results in Figure 1, we are led to the conclusion that the pH-dependent change in HER kinetics at Au(111) in alkaline media are due to changes in the OH⁻ concentration. To explain this, we note that OH⁻ has multiple roles in the HER. Thermodynamically, higher bulk OH⁻ concentrations shift the equilibrium potential of HER towards negative potentials (via the Nernst equation: $E_{\text{eq,SHE}} = E^0_{\text{eq,SHE}} + \frac{RT}{2F}\ln\frac{a^b_{H_2O}}{a^b_{H_2}a^b_{OH^-}}$), which makes HER thermodynamically less favorable under higher OH⁻ concentrations at a fixed electrode potential on the SHE scale. On the other hand, mechanistically OH⁻ is a product of alkaline HER and needs to diffuse from the EDL to the bulk solution to complete the reaction cycle. Thus, efficient hydrogen-bond connectivity among interfacial water molecules and OH⁻ is critical. The presence of OH⁻ may also change the reaction kinetics by controlling the hydrogen bond network and by coordinating with the reacting water molecule. We propose that OH⁻ may enhance, rather than disrupt the hydrogen bond network, and consequently reduce the barrier for the Volmer reaction at Au(111). When these options are considered, Eq. (2) is modified to explicitly account for changes in the Volmer barrier:

$$\Delta G^{\neq}_{\text{alkaV}}(E_{\text{SHE}}) = \Delta G^{\neq,0}_{\text{alkaV}} + \alpha_{\text{alkaV}}F(E_{\text{SHE}} - \phi^{\text{RP}}_{\text{alka}} - E^0_{\text{SHE,alkaV}}) + \Delta\Delta G^{\neq}_{\text{alkaV}} \qquad (3)$$

where $\Delta\Delta G^{\neq}_{\text{alkaV}}$ includes all pH-dependent changes in the activation energy of the alkaline Volmer reaction resulting from changing the bulk OH⁻ concentration. This term uses pH=10 as a reference ($\Delta\Delta G^{\neq}_{\text{alkaV}} = 0$ eV at pH=10). By solving the modified Poisson-Nernst-Planck equation [47] and coupling Eqs (1) and (3), with the equations and parameters used in the EDL model detailed in the SI, we obtain the results shown Figure 3. By fixing all parameters apart from $\Delta\Delta G^{\neq}_{\text{alkaV}}$, an excellent agreement between the measured and simulated polarization curves is obtained. We find that $\Delta\Delta G^{\neq}_{\text{alkaV}}$ is 0 eV, -0.013 eV and -0.02 eV, respectively at pH=10, 11, 12 and 13. These results show that the barrier for the Volmer reaction does indeed decrease as the OH⁻ concentration increases. This implies that further increase of OH⁻ concentration may further boost the HER .

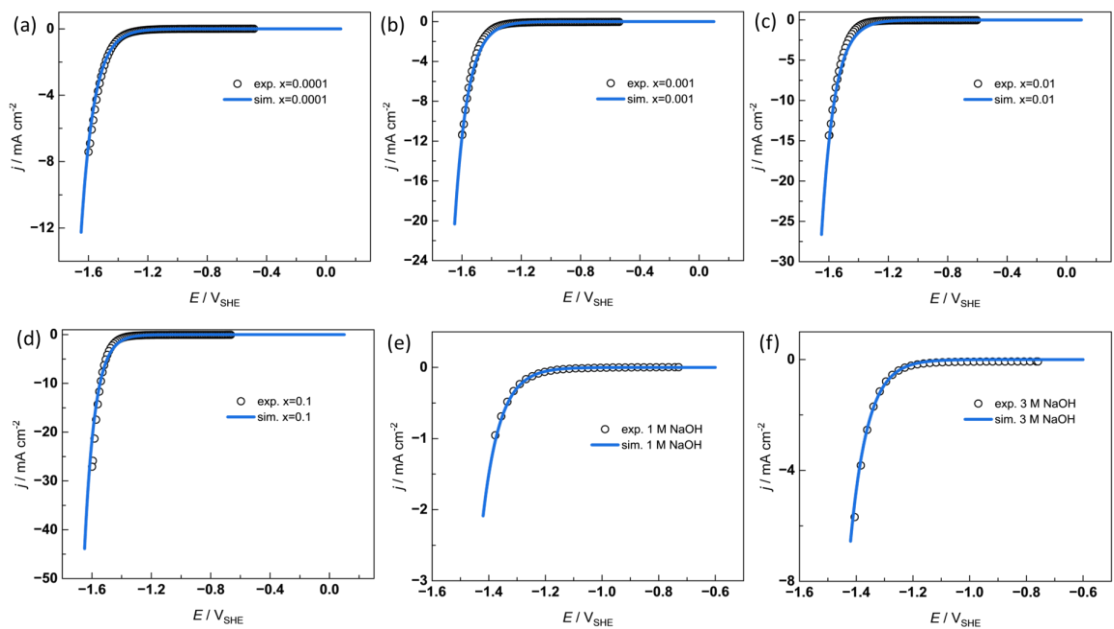

Figure 3 Polarization curves from experiments (solid line) and the kinetic + EDL model (circle) for HER at Au(111) surface (a-d) x M NaOH+(0.1-x) M NaClO$_4$ (x =0.0001,0.001,0.01, 0.1) and (e) 1 M and (f) 3 M NaOH solutions on SHE scale. Some important parameters are given here. $\Delta G_{\text{alkaV}}^{\neq,0} = 0.624$ eV @ pH = 10, $\alpha_{\text{alkaV}} = 0.4$, $E_{\text{SHE,alkaV}}^0 = -1.24$ eV.

To validate this hypothesis, we extended the considered pH range and HER experiments were conducted in 1–9 M NaOH solutions. As shown in Figures 4a and 4b, the HER currents at identical potentials (*vs*. RHE or SHE) increase significantly as the OH⁻ concentration is increased. Chronoamperometric measurements at −0.4 V$_{\text{RHE}}$ and −0.5 V$_{\text{RHE}}$ further confirm this trend (Figure 4c). To rule out the possibility that the pH changes the surface morphology or structure, basic cyclic voltammograms (CVs) were recorded in $x$ M NaOH solution (Figure 4d) and 0.05 M H$_2$SO$_4$ solution (Figure 4e) immediately right after recording the HER polarization curves. These results confirm that the Au(111) electrode remained stable and unchanged under all considered conditions. Figure 4f displays the basic CVs of Au(111) measured in 0.05 M H$_2$SO$_4$ solution after recording the $j-t$ curves for HER at -0.5 V$_{\text{RHE}}$ for 5 minutes and then removed from the solution and rinsed thoroughly with water. The features of Figure 4f closely match those observed in Figure 4e. Simulations of the $j-E$ curves obtained with 1 M (Figure 3e) and 3 M (Figure 3f) NaOH show further decrease of the $\Delta\Delta G_{\text{alkaV}}^{\neq}$ to -0.056 eV and -0.066 eV, respectively. The enhancement of HER current by the increased Na⁺ concentration can be excluded since Figure S3 shows that the addition of NaCl while keeping the OH⁻ concentration constant (0.1 M) leads to an obvious drop in the HER current at Au(111). This is in agreement with previous finding by Koper et

al.[40]

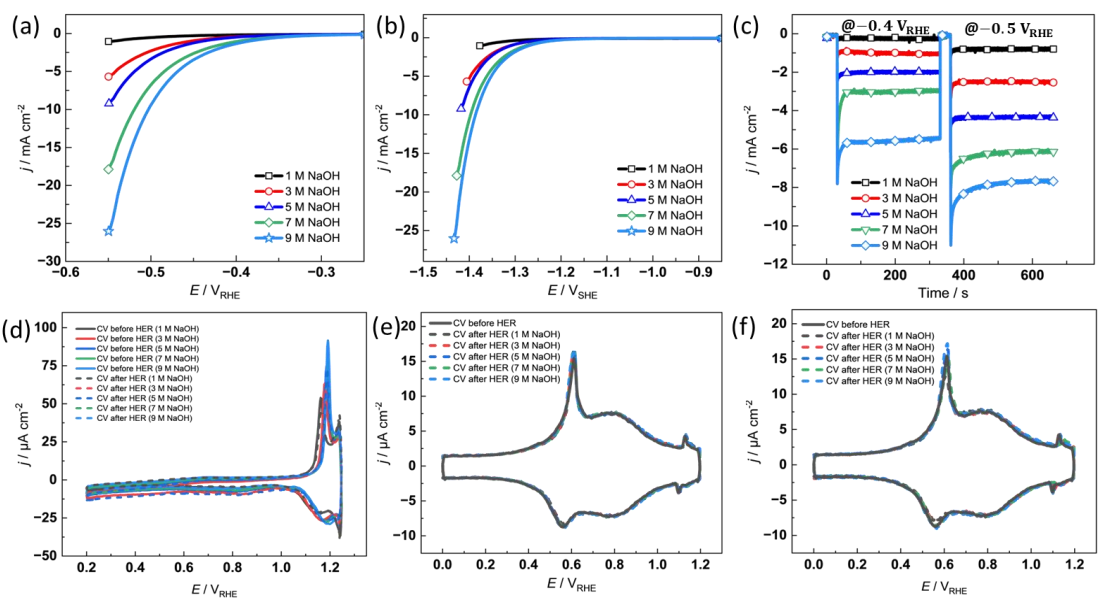

Figure 4 Polarization curves (a, b) and current transients ($j-t$ curves) recorded at -0.4 $V_{RHE}$ and -0.5 $V_{RHE}$ for HER at Au(111) electrode in Ar-saturated x M NaOH (x=1,3,5,7,9) solutions on the (a) RHE scale and (b) SHE scale, scan rate: at 10 mV/s, electrode rotation speed: 2500 rpm. All potentials were corrected for the iR-drop effect. Basic CVs were recorded in $x$ M NaOH solution (d) and 0.05 M $H_2SO_4$ solution (e) immediately after measuring the $j-E$ curves. (f) Basic CVs in 0.05 M $H_2SO_4$ after current transients ($j-t$ curves) measured at -0.5 $V_{RHE}$ for 5 min. All basic CVs are recorded with a potential scan rate of 50 mV/s.

We further used in situ infrared (IR) spectroscopy to address how and why the interfacial $OH^-$ concentration affects the solvent properties and the microscopic mechanism of alkaline HER. As shown in Figure 5a, the O-H stretching vibration frequency of water in concentrated NaOH solutions displays a significant redshift below 3000 cm$^{-1}$. A similar decrease in the O-H stretching vibration frequency is observed at the Au/x M NaOH interface (Figure 5b-c), where IR spectra of interfacial water and $OH^-$ are recorded under attenuated total reflection configuration mode. Figures 5b-c show that the O-H stretching frequency shifts to lower values as the $OH^-$ concentration increases and when the potential is decreased; these results indicate that higher $OH^-$ concentrations and more reducing potentials weaken the O-H bond of water.[55] As a decrease in the O-H vibrational frequency is also known to correspond to stronger hydrogen bond between water molecules,[56] our IR results also indicate that the presence of $OH^-$ and reducing potentials strengthen the hydrogen bonding network. In addition to changes in the characteristic bending and stretching vibrations of water molecules at 1600-1700 cm$^{-1}$ and 3100-3400 cm$^{-1}$, respectively, two new features are observed as the OH⁻ concentration increases and as the potential is made more reducing. The new shoulder peak around 3600 cm⁻¹ is due to the OH⁻ itself while the broad peak between 1200-3000 cm⁻¹ arises from the H-O-H···OH⁻ interaction. The latter interaction leads to modifications in the hydrogen bonding network, which is expected to be strengthened as witnessed by the decreased O-H vibrational frequency discussed earlier. Overall, the IR results indicate that OH⁻ anions and reductive potentials lead to pronounced changes in the interfacial hydrogen bond network. We attribute these changes to two synergistic effects: (1) direct coordination between OH⁻ and H₂O which weakens the O-H bond of water molecules to be broken in the Volmer step and (2) the strengthened hydrogen-bond network between the water molecules and/or water molecules and OH⁻.

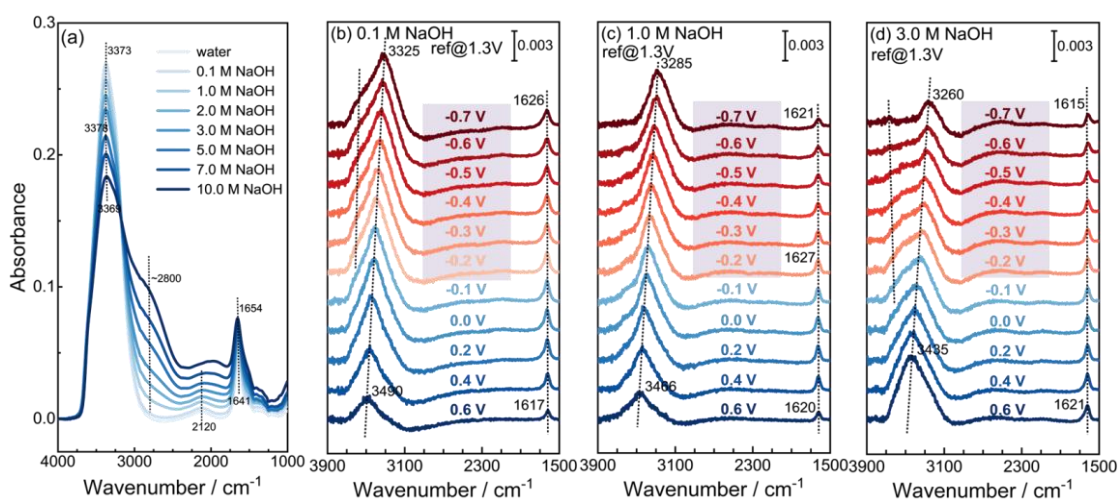

Figure 5: (a) Absorption spectra of NaOH solutions of different concentrations recorded in the transmission mode. (b-d) ATR-FTIR spectra of Au thin film electrode / x M NaOH electrolyte solution interface, x=0.1, 1, 3 M. The bending vibration frequency redshifts at more negative potentials, this does not support the HBN is strengthened.

Overall, our experiments show that the OH⁻ concentration plays a pivotal role in the HER activity and kinetics. To provide mechanistic understanding of the OH anion's role in HER, we recap our main findings: 1) For the case with pH=10 to 13, the HER activity increases with pH when the Na⁺ concentration is kept constant, switching the supporting electrolyte from ClO₄⁻ to Cl⁻ does not influence the HER kinetics; 2) at pH≥13, when keeping the OH⁻ concentration constant, addition of Na⁺ concentration leads to a significant drop in the HER current; 3) the vibrational spectra shows clear signatures of weakened O-H bond vibration at negative potentials as the intensity of such vibrational peaks increases with OH⁻ concentration and the negative shift of the electrode potential.

Based on these findings, we propose a new mechanistic framework to explain the OH anion's dual role in alkaline HER. Because the potential of surface free charge does not depend on pH, at constant bulk Na$^+$ concentration and increasing pH (Figure. 6a and 6b), the interfacial cation concentration remains unchanged. The gradual enrichment of OH$^-$ beyond the EDL moderately changes the hydrogen-bond network and facilitates the lowering of the energy barrier for proton transfer from bulk solution to the electrode surface, thereby enhancing HER activity. When increasing NaOH concentration from 1 to 9 M (Figure. 6c), the Debye length calculated through Gouy-Chapman-Stern theory contracts from 0.3 nm to 0.1 nm[57] and this EDL compression leads to nearly complete potential drop within the compact layer and eliminates the potential gradient between the diffuse layer and bulk solution. Consequently, the interfacial Na$^+$ and OH$^-$ concentrations approach bulk levels (1-9 M) while moving closer to the electrode surface. Excessive surface Na$^+$ ions form obstructive hydrated clusters that disrupt hydrogen-bond connectivity in the EDL, inhibiting free water molecules from reaching the electrode surface. However, the interfacial OH$^-$ enrichment counteracts this limitation through two synergistic effects: First, it reconstructs the proton transport channels and water structure by strengthening the OH$^-$-H$_2$O hydrogen bonds; Second, it elongates and weakens water's O-H bonds, and thereby reduces the activation energy of the Volmer reaction.

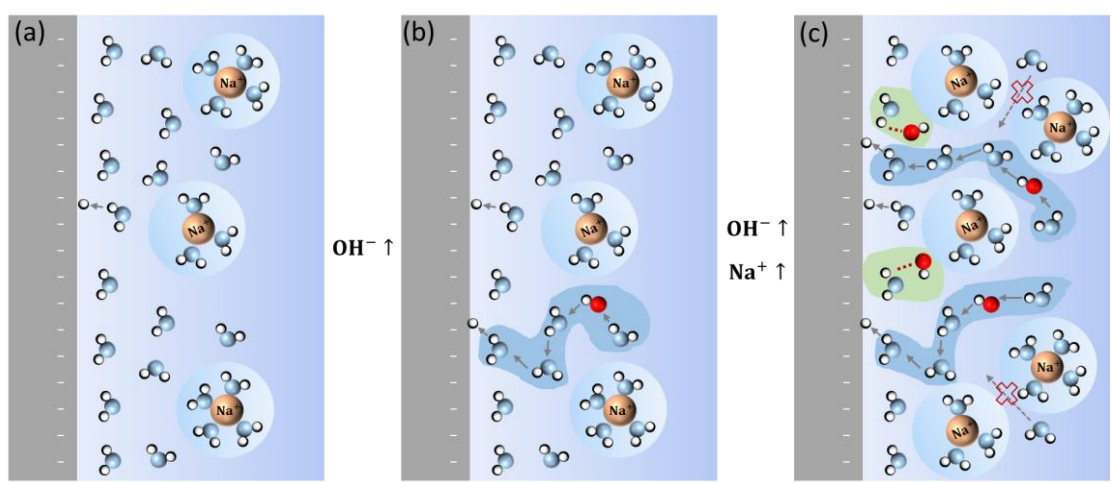

Figure 6. Mechanistic schematics of Na$^+$ and OH$^-$ interactions at the Au(111) electrode surface. (a)→(b) Increasing the bulk OH$^-$ concentration while fixing the bulk concentration of Na$^+$ leads to an increase in the interfacial OH$^-$ concentration while the interfacial cation concentration is unchanged. The increased HER activity is due to the enhanced hydrogen transport, strengthened hydrogen bonding network, and weakened O-H bond strength; (b)→(c) Simultaneous increase in both bulk concentrations of Na$^+$ and OH$^-$ changes the interfacial cation and anion concentration.

These two effects counteract: $Na^+$ decreases the hydrogen bond connectivity and inhibits water transport while $OH^-$ enhances both. The latter effect is more dominant and leads to improved HER kinetics at higher $OH^-$ concentrations.

## 4. Conclusion

This work elucidates the pivotal role of $OH^-$ in enhancing alkaline HER kinetics at Au(111). By decoupling the pH and cation effects, we reveal that increasing $OH^-$ concentration—even under ultrahigh NaOH conditions—continuously boosts the HER activity by outperforming the predicted inhibition from cation crowding. The origin of the $OH^-$-induced enhancement in the alkaline HER kinetics is explained by the hydroxide anion's capability to 1) weaken the O–H bonds in water molecules, which consequently reduces the Volmer step barrier, and 2) facilitate proton transfer through strengthening the hydrogen bonding network. Notably, $OH^-$ compensates for cation-induced steric limitations by optimizing the interfacial water structure and hydrogen-bond connectivity. These insights challenge the traditional cation-centric view where cations are the only electrolyte species that can control the alkaline HER performance. Instead, we show that both the cation and anion effects can be understood through a unified mechanism where structure-making ions such as $Li^+$, $K^+$, and $OH^-$ enhance alkaline HER kinetics by strengthening the hydrogen bond network and its connectivity at the electrochemical interface. Our findings provide a mechanistic foundation for advancing green hydrogen technologies, emphasizing the synergy between electrolyte composition and catalyst-electrolyte interfaces in governing reaction dynamics.

**Acknowledgements**

This work is supported by National Key Research and Development Program of China (No. 2023YFA1509004) and Chinese national science foundation (No. 22172151). M.M.M acknowledges funding by the Academy of Finland (grant #338228).